\documentclass[11pt]{article}
\usepackage{times}
\usepackage[margin=2cm]{geometry}
\usepackage{graphicx}
\usepackage{amsmath,amssymb,amstext}

\begin{document}

  \title{On exact relations for the calculation of effective properties of composites}

  \author{Biswajit Banerjee (b.banerjee.nz@gmail.com) and Daniel O. Adams \\
          Dept. of Mechanical Engineering, University of Utah, Salt Lake City, USA.}
  \date{June 2002}

  %\address{
  %	   50 S Central Campus Drive, Salt Lake City, UT 84112, USA,
  %           Fax: (801)-585-9826.}
  %\eads{\mailto{banerjee@eng.utah.edu},\mailto{adams@mech.utah.edu}}

  \maketitle

  \section*{Abstract}
  {\small
    Numerous exact relations exist that relate the effective elastic properties of composites to the elastic properties of their components.  These relations can not only be used to determine the properties of certain composites, but also provide checks on the accuracy on numerical techniques for the calculation of effective properties.  In this work, some exact relations are discussed and estimates from finite element calculations, the generalized method of cells and the recursive cell method are compared with estimates from the exact relations.  Comparisons with effective properties predicted using exact relations show that the best estimates are obtained from the finite element calculations while the moduli are overestimated by the recursive cell method and underestimated by the generalized method of cells.  However, not all exact relations can be used to make such a distinction.

  }

  %
%  Compositions of PBX components
%
  %\def\RDX{{\footnotesize RDX : C$_3$ H$_6$ N$_6$ O$_6$}}
  %\def\TATB{{\footnotesize TATB : triaminotrinitrobenzene}}
  %\def\Viton{{\footnotesize Viton : random copolymer of hexfluoropropane
%			     and vinylidene fluoride (1:2)}}
  %\def\KE{{\footnotesize KEL-F-800 : random copolymer of 
%		             chlorotrifluoroethylene
%			     and vinylidene fluoride (3:1)}}
  %\def\KS{{\footnotesize KEL-F-3700 : (CFClCF$_2$CH$_2$CF$_2$)$_n$}}
  %\def\Exon{{\footnotesize Exon-461 : (CFClCF$_2$CH$_2$CF$_2$)$_n$}}
  %\def\HTPB{{\footnotesize HTPB : hydroxyl terminated poly butadiene}}
  %\def\NC{{\footnotesize NC : nitrocellulose}}
  %\def\CEF{{\footnotesize CEF : chloroethyl phosphate}}
  %\def\KT{{\footnotesize K10 : plasticizer (composition not known) }}
  %\def\HMX{{\footnotesize HMX : 1,3,5,7-tetranitro-1,3,5,7-tetraazacyclooctane}}
  %\def\Estane{{\footnotesize Estane 5703 : segmented polyeurethene of low
%	molecular weight poly(butylene adipate) soft 
%	segments and 4,4 diphenylmethanediisocyanate-1,4- 
%	butanediol hard segments.}}
  %\def\BDN{{\footnotesize BDNPA/F : bis-dinitropropylacetal/formal}}

  \def\RDX{{RDX : C$_3$ H$_6$ N$_6$ O$_6$}}
  \def\TATB{{TATB : triaminotrinitrobenzene}}
  \def\Viton{{Viton : random copolymer of hexfluoropropane
			     and vinylidene fluoride (1:2)}}
  \def\KE{{KEL-F-800 : random copolymer of 
		             chlorotrifluoroethylene
			     and vinylidene fluoride (3:1)}}
  \def\KS{{KEL-F-3700 : (CFClCF$_2$CH$_2$CF$_2$)$_n$}}
  \def\Exon{{Exon-461 : (CFClCF$_2$CH$_2$CF$_2$)$_n$}}
  \def\HTPB{{HTPB : hydroxyl terminated poly butadiene}}
  \def\NC{{NC : nitrocellulose}}
  \def\CEF{{CEF : chloroethyl phosphate}}
  \def\KT{{K10 : plasticizer (composition not known) }}
  \def\HMX{{HMX : 1,3,5,7-tetranitro-1,3,5,7-tetraazacyclooctane}}
  \def\Estane{{Estane 5703 : segmented polyeurethene of low
	molecular weight poly(butylene adipate) soft 
	segments and 4,4 diphenylmethane diisocyanate 1,4 
	butanediol hard segments.}}
  \def\BDN{{BDNPA/F : bis-dinitropropylacetal/formal}}
\newcommand{\Pin}{\parindent 1.5em}
\newcommand{\D}{\displaystyle}
\newcommand{\Scs}{\scriptstyle}
\newcommand{\Te}{\text{eff}}
\newcommand{\Tv}{\text{V}}
\newcommand{\Ts}{\text{S}}
\newcommand{\TwoD}{\text{2D}}
\newcommand{\Tta}{\text{a}}
\newcommand{\Ttb}{\text{b}}
\newcommand{\Ttc}{\text{c}}
\newcommand{\Ttd}{\text{d}}
\newcommand{\Tte}{\text{e}}
\newcommand{\Ttf}{\text{f}}
\newcommand{\Ttg}{\text{g}}
\newcommand{\Tth}{\text{h}}
\newcommand{\Tti}{\text{i}}
\newcommand{\Ttj}{\text{j}}
\newcommand{\Ttk}{\text{k}}
\newcommand{\Ttl}{\text{l}}
\newcommand{\Ttm}{\text{m}}
\newcommand{\Ttn}{\text{n}}
\newcommand{\Tto}{\text{o}}
\newcommand{\Ttp}{\text{p}}
\newcommand{\Ttq}{\text{q}}
\newcommand{\Ttr}{\text{r}}
\newcommand{\Tts}{\text{s}}
\newcommand{\Over}[1]{\ensuremath{\cfrac{1}{#1}}}
\newcommand{\Avg}[1]{\ensuremath{\left<#1\right>}}
\newcommand{\VAvg}[1]{\ensuremath{\Avg{#1}_{\Tv}}}
\newcommand{\SAvg}[1]{\ensuremath{\Avg{#1}_{\Ts}}}
\newcommand{\Ceff}[1]{\ensuremath{C^{\Te}_{#1}}}
\newcommand{\Seff}[1]{\ensuremath{S^{\Te}_{#1}}}
\newcommand{\Eeff}[1]{\ensuremath{E^{\Te}_{#1}}}
\newcommand{\Neff}[1]{\ensuremath{\nu^{\Te}_{#1}}}
\newcommand{\Geff}[1]{\ensuremath{G^{\Te}_{#1}}}
\newcommand{\Sigx}{\sigma_{11}}
\newcommand{\Sigy}{\sigma_{22}}
\newcommand{\Sigz}{\sigma_{33}}
\newcommand{\Tauxy}{\sigma_{12}}
\newcommand{\Tauyz}{\sigma_{23}}
\newcommand{\Tauzx}{\sigma_{31}}
\newcommand{\Sigxy}{\sigma_{12}}
\newcommand{\Sigyz}{\sigma_{23}}
\newcommand{\Sigzx}{\sigma_{31}}
\newcommand{\Epsx}{\epsilon_{11}}
\newcommand{\Epsy}{\epsilon_{22}}
\newcommand{\Epsz}{\epsilon_{33}}
\newcommand{\Gamxy}{\epsilon_{12}}
\newcommand{\Gamyz}{\epsilon_{23}}
\newcommand{\Gamzx}{\epsilon_{31}}
\newcommand{\ASigx}{\VAvg{\Sigx}}
\newcommand{\ASigy}{\VAvg{\Sigy}}
\newcommand{\ASigz}{\VAvg{\Sigz}}
\newcommand{\ATauxy}{\VAvg{\Tauxy}}
\newcommand{\ATauyz}{\VAvg{\Tauyz}}
\newcommand{\ATauzx}{\VAvg{\Tauzx}}
\newcommand{\ASigxy}{\VAvg{\Sigxy}}
\newcommand{\ASigyz}{\VAvg{\Sigyz}}
\newcommand{\ASigzx}{\VAvg{\Sigzx}}
\newcommand{\AEpsx}{\VAvg{\Epsx}}
\newcommand{\AEpsy}{\VAvg{\Epsy}}
\newcommand{\AEpsz}{\VAvg{\Epsz}}
\newcommand{\AGamxy}{\VAvg{\Gamxy}}
\newcommand{\AGamyz}{\VAvg{\Gamyz}}
\newcommand{\AGamzx}{\VAvg{\Gamzx}}
\newcommand{\Ca}{\Ceff{11}}
\newcommand{\Cb}{\Ceff{12}}
\newcommand{\Cc}{\Ceff{66}}
\newcommand{\Cd}{\Ceff{22}}
%
% GMC notation
%
\newcommand{\Xx}{\ensuremath{x^{(\alpha)}_{1}}}
\newcommand{\Yy}{\ensuremath{x^{(\beta)}_{2}}}
\newcommand{\Zz}{\ensuremath{x^{(\gamma)}_{3}}}
\newcommand{\Xxh}{\ensuremath{x^{(\hat\alpha)}_{1}}}
\newcommand{\Yyh}{\ensuremath{x^{(\hat\beta)}_{2}}}
\newcommand{\Zzh}{\ensuremath{x^{(\hat\gamma)}_{3}}}
\newcommand{\Ddx}{\ensuremath{\frac{\partial}{\partial X_1}}}
\newcommand{\Ddy}{\ensuremath{\frac{\partial}{\partial X_2}}}
\newcommand{\Ddz}{\ensuremath{\frac{\partial}{\partial X_3}}}
\newcommand{\LocXYZ}{\Xx,\Yy,\Zz}
\newcommand{\GloXYZ}{\ensuremath{X_1,X_2,X_3}}
\newcommand{\Vabg}
   {\ensuremath{\frac{1}{8h^3}\int_{-h}^h\int^h_{-h}\int_{-h}^h} a~d\Xx d\Yy d\Zz }
\newcommand{\Vint}{\ensuremath{\frac{1}{8h^3}\int_{V_s}}}
\newcommand{\Vdsum} {\ensuremath{\frac{1}{8N^3h^3}\sum^{N}_{\alpha = 1}
	\sum^{N}_{\beta = 1}\sum^{N}_{\gamma = 1}}}
\newcommand{\Vsum}{\ensuremath{\frac{1}{V}\sum^N}}
\newcommand{\Aintabg}{\ensuremath{\int^{h}_{-h} \int^{h}_{-h}}}
\newcommand{\Aint}{\ensuremath{\int_{A_I}}}
\newcommand{\Abg}{\ensuremath{(\alpha \beta \gamma)}}
\newcommand{\Ahbg}{\ensuremath{(\hat{\alpha} \beta \gamma)}}
\newcommand{\Abhg}{\ensuremath{(\alpha \hat{\beta} \gamma)}}
\newcommand{\Abgh}{\ensuremath{(\alpha \beta \hat{\gamma})}}
\newcommand{\Da}{\ensuremath{d_ {\alpha}}}
\newcommand{\Hb}{\ensuremath{h_ {\beta}}}
\newcommand{\Lg}{\ensuremath{l_ {\gamma}}}
\newcommand{\Dah}{\ensuremath{d_ {\hat\alpha}}}
\newcommand{\Hbh}{\ensuremath{h_ {\hat\beta}}}
\newcommand{\Lgh}{\ensuremath{l_ {\hat\gamma}}}
\newcommand{\Epsxxs}{\SAvg{\epsilon^{\Abg}_{11}}}
\newcommand{\Epsxys}{\SAvg{\epsilon^{\Abg}_{12}}}
\newcommand{\Epsxzs}{\SAvg{\epsilon^{\Abg}_{13}}}
\newcommand{\Epsyxs}{\SAvg{\epsilon^{\Abg}_{21}}}
\newcommand{\Epsyys}{\SAvg{\epsilon^{\Abg}_{22}}}
\newcommand{\Epsyzs}{\SAvg{\epsilon^{\Abg}_{23}}}
\newcommand{\Epszxs}{\SAvg{\epsilon^{\Abg}_{31}}}
\newcommand{\Epszys}{\SAvg{\epsilon^{\Abg}_{32}}}
\newcommand{\Epszzs}{\SAvg{\epsilon^{\Abg}_{33}}}
\newcommand{\Epsxy}{\VAvg{\epsilon_{12}}}
\newcommand{\Epsyz}{\VAvg{\epsilon_{23}}}
\newcommand{\Epszx}{\VAvg{\epsilon_{31}}}
\newcommand{\Epsxz}{\VAvg{\epsilon_{13}}}
\newcommand{\RR}[1]{\ensuremath{R^{\Abg}_{#1}}}
\newcommand{\TT}[1]{\ensuremath{\tau^{\Abg}_{#1}}}
\newcommand{\Rx}{\ensuremath{R^{\Abg}_{1i}}}
\newcommand{\Ry}{\ensuremath{R^{\Abg}_{2i}}}
\newcommand{\Rz}{\ensuremath{R^{\Abg}_{3i}}}
\newcommand{\Rij}{\ensuremath{R^{\Abg}_{ij}}}
\newcommand{\Rji}{\ensuremath{R^{\Abg}_{ji}}}
\newcommand{\Taux}{\ensuremath{\tau^{\Abg}_{1i}}}
\newcommand{\Tauy}{\ensuremath{\tau^{\Abg}_{2i}}}
\newcommand{\Tauz}{\ensuremath{\tau^{\Abg}_{3i}}}
\newcommand{\Tauij}{\ensuremath{\tau^{\Abg}_{ij}}}
\newcommand{\Tauji}{\ensuremath{\tau^{\Abg}_{ji}}}

\newcommand{\bA}{\ensuremath{\mathbf{A}}}
\newcommand{\Sx}{\ensuremath{\left<\sigma_{x}\right>}}
\newcommand{\Sy}{\ensuremath{\left<\sigma_{y}\right>}}
\newcommand{\Ex}{\ensuremath{\left<\epsilon_{x}\right>}}
\newcommand{\Ey}{\ensuremath{\left<\epsilon_{y}\right>}}
\newcommand{\Exy}{\ensuremath{\left<\epsilon_{xy}\right>}}
\newcommand{\Exp}[1]{\ensuremath{\times 10^{#1}}}
\newcommand{\Eff}[2]{\ensuremath{{#1}^{\Te}_{#2}}}
\newcommand{\Ti}[1]{\ensuremath{(\times 10^{#1})}}
\newcommand{\Sigmaij}{\ensuremath{\left< \sigma_{ij} \right>}}
\newcommand{\SigmaxRVE}{\ensuremath{\left< \sigma_{11} \right>}}
\newcommand{\SigmayRVE}{\ensuremath{\left< \sigma_{22} \right>}}
\newcommand{\SigmazRVE}{\ensuremath{\left< \sigma_{33} \right>}}
\newcommand{\SigmaxyRVE}{\ensuremath{\left< \sigma_{12} \right>}}
\newcommand{\SigmayzRVE}{\ensuremath{\left< \sigma_{23} \right>}}
\newcommand{\SigmaxzRVE}{\ensuremath{\left< \sigma_{13} \right>}}
\newcommand{\Sigmaijs}{\ensuremath{\left< \sigma^{\Abg}_{ij} \right>}}
\newcommand{\Sigmax}{\ensuremath{\left< \sigma^{\Abg}_{1i} \right>}}
\newcommand{\Sigmaxx}{\ensuremath{\left< \sigma^{\Abg}_{11} \right>}}
\newcommand{\Sigmaxy}{\ensuremath{\left< \sigma^{\Abg}_{12} \right>}}
\newcommand{\Sigmaxz}{\ensuremath{\left< \sigma^{\Abg}_{13} \right>}}
\newcommand{\Sigmay}{\ensuremath{\left< \sigma^{\Abg}_{2i} \right>}}
\newcommand{\Sigmayx}{\ensuremath{\left< \sigma^{\Abg}_{21} \right>}}
\newcommand{\Sigmayy}{\ensuremath{\left< \sigma^{\Abg}_{22} \right>}}
\newcommand{\Sigmayz}{\ensuremath{\left< \sigma^{\Abg}_{23} \right>}}
\newcommand{\Sigmaz}{\ensuremath{\left< \sigma^{\Abg}_{3i} \right>}}
\newcommand{\Sigmazx}{\ensuremath{\left< \sigma^{\Abg}_{31} \right>}}
\newcommand{\Sigmazy}{\ensuremath{\left< \sigma^{\Abg}_{32} \right>}}
\newcommand{\Sigmazz}{\ensuremath{\left< \sigma^{\Abg}_{33} \right>}}
\newcommand{\Sigmaxh}{\ensuremath{\left< \sigma^{\Ahbg}_{1i} \right>}}
\newcommand{\Sigmaxxh}{\ensuremath{\left< \sigma^{\Ahbg}_{11} \right>}}
\newcommand{\Sigmaxyh}{\ensuremath{\left< \sigma^{\Ahbg}_{12} \right>}}
\newcommand{\Sigmaxzh}{\ensuremath{\left< \sigma^{\Ahbg}_{13} \right>}}
\newcommand{\Sigmayh}{\ensuremath{\left< \sigma^{\Abhg}_{2i} \right>}}
\newcommand{\Sigmayxh}{\ensuremath{\left< \sigma^{\Abhg}_{21} \right>}}
\newcommand{\Sigmayyh}{\ensuremath{\left< \sigma^{\Abhg}_{22} \right>}}
\newcommand{\Sigmayzh}{\ensuremath{\left< \sigma^{\Abhg}_{23} \right>}}
\newcommand{\Sigmazh}{\ensuremath{\left< \sigma^{\Abgh}_{3i} \right>}}
\newcommand{\Sigmazxh}{\ensuremath{\left< \sigma^{\Abgh}_{31} \right>}}
\newcommand{\Sigmazyh}{\ensuremath{\left< \sigma^{\Abgh}_{32} \right>}}
\newcommand{\Sigmazzh}{\ensuremath{\left< \sigma^{\Abgh}_{33} \right>}}
\newcommand{\Suma}{\ensuremath{\sum^{N}_{\alpha=1}}}
\newcommand{\Sumb}{\ensuremath{\sum^{N}_{\beta=1}}}
\newcommand{\Sumg}{\ensuremath{\sum^{N}_{\gamma=1}}}
\newcommand{\Sumai}{\ensuremath{\sum^{N}_{i=1}}}
\newcommand{\Sumaj}{\ensuremath{\sum^{N}_{j=1}}}
\newcommand{\Sumbi}{\ensuremath{\sum^{N}_{i=1}}}
\newcommand{\Sumbj}{\ensuremath{\sum^{N}_{j=1}}}
\newcommand{\Sumgi}{\ensuremath{\sum^{N}_{i=1}}}
\newcommand{\Sumgj}{\ensuremath{\sum^{N}_{j=1}}}
\newcommand{\Epspx}{\ensuremath{\left< \epsilon^{p}_{11} \right>}}
\newcommand{\Epspy}{\ensuremath{\left< \epsilon^{p}_{22} \right>}}
\newcommand{\Epspz}{\ensuremath{\left< \epsilon^{p}_{33} \right>}}
\newcommand{\Epspxy}{\ensuremath{\left< \epsilon^{p}_{12} \right>}}
\newcommand{\Epspyz}{\ensuremath{\left< \epsilon^{p}_{23} \right>}}
\newcommand{\Epspxz}{\ensuremath{\left< \epsilon^{p}_{13} \right>}}
\newcommand{\Inta}{\ensuremath{\int^{\Da}_{-\Da}}}
\newcommand{\Intb}{\ensuremath{\int^{\Hb}_{-\Hb}}}
\newcommand{\Intg}{\ensuremath{\int^{\Lg}_{-\Lg}}}
\newcommand{\Phix}{\ensuremath{\Phi^{\Abg}_{1}}}
\newcommand{\Phiy}{\ensuremath{\Phi^{\Abg}_{2}}}
\newcommand{\Phiz}{\ensuremath{\Phi^{\Abg}_{3}}}
\newcommand{\Chix}{\ensuremath{\Theta^{\Abg}_{1}}}
\newcommand{\Chiy}{\ensuremath{\Theta^{\Abg}_{2}}}
\newcommand{\Chiz}{\ensuremath{\Theta^{\Abg}_{3}}}
\newcommand{\Psix}{\ensuremath{\Psi^{\Abg}_{1}}}
\newcommand{\Psiy}{\ensuremath{\Psi^{\Abg}_{2}}}
\newcommand{\Psiz}{\ensuremath{\Psi^{\Abg}_{3}}}
\newcommand{\Epsmat}{\ensuremath{\left<\boldsymbol{\epsilon}^{\Abg}\right>}}
\newcommand{\Epspmat}{\ensuremath{\left<\boldsymbol{\epsilon}^{p\Abg}\right>}}
\newcommand{\Epstmat}{\ensuremath{\boldsymbol{\alpha}^{\Abg}}}
\newcommand{\Sigmamat}{\ensuremath{\left<\boldsymbol{\sigma}^{\Abg}\right>}}
\newcommand{\Smat}{\ensuremath{\boldsymbol{S}^{\Abg}}}
\newcommand{\Sxx}{\ensuremath{S^{\Abg}_{11}}}
\newcommand{\Sxy}{\ensuremath{S^{\Abg}_{12}}}
\newcommand{\Sxz}{\ensuremath{S^{\Abg}_{13}}}
\newcommand{\Syx}{\ensuremath{S^{\Abg}_{21}}}
\newcommand{\Syy}{\ensuremath{S^{\Abg}_{22}}}
\newcommand{\Syz}{\ensuremath{S^{\Abg}_{23}}}
\newcommand{\Szx}{\ensuremath{S^{\Abg}_{31}}}
\newcommand{\Szy}{\ensuremath{S^{\Abg}_{32}}}
\newcommand{\Szz}{\ensuremath{S^{\Abg}_{33}}}
\newcommand{\Sxyxy}{\ensuremath{S^{\Abg}_{66}}}
\newcommand{\Syzyz}{\ensuremath{S^{\Abg}_{44}}}
\newcommand{\Sxzxz}{\ensuremath{S^{\Abg}_{55}}}
\newcommand{\Epstx}{\ensuremath{\alpha^{\Abg}_{11}}}
\newcommand{\Epsty}{\ensuremath{\alpha^{\Abg}_{22}}}
\newcommand{\Epstz}{\ensuremath{\alpha^{\Abg}_{33}}}
\newcommand{\Txx}{\ensuremath{T^{(\beta\gamma)}_{11}}}
\newcommand{\Tyy}{\ensuremath{T^{(\alpha\gamma)}_{22}}}
\newcommand{\Tzz}{\ensuremath{T^{(\alpha\beta)}_{33}}}
\newcommand{\Txy}{\ensuremath{T^{(\beta\gamma)}_{12}}}
\newcommand{\Tyx}{\ensuremath{T^{(\alpha\gamma)}_{21}}}
\newcommand{\TTxy}{\ensuremath{T^{(\gamma)}_{12}}}
\newcommand{\Tyz}{\ensuremath{T^{(\alpha\gamma)}_{23}}}
\newcommand{\Tzy}{\ensuremath{T^{(\alpha\beta)}_{32}}}
\newcommand{\TTyz}{\ensuremath{T^{(\alpha)}_{23}}}
\newcommand{\Txz}{\ensuremath{T^{(\beta\gamma)}_{13}}}
\newcommand{\Tzx}{\ensuremath{T^{(\alpha\beta)}_{31}}}
\newcommand{\TTxz}{\ensuremath{T^{(\beta)}_{13}}}
\newcommand{\Na}{\ensuremath{N}}
\newcommand{\Nb}{\ensuremath{N}}
\newcommand{\Ng}{\ensuremath{N}}
\newcommand{\Nabg}{\ensuremath{N_{\alpha\beta\gamma}}}
\newcommand{\Nab}{\ensuremath{N_{\alpha\beta}}}
\newcommand{\Nbg}{\ensuremath{N_{\beta\gamma}}}
\newcommand{\Alx}{\ensuremath{\alpha^{\Abg}_{11}}}
\newcommand{\Aly}{\ensuremath{\alpha^{\Abg}_{22}}}
\newcommand{\Alz}{\ensuremath{\alpha^{\Abg}_{33}}}
\newcommand{\Alxstar}{\ensuremath{\alpha^{\Te}_{11}}}
\newcommand{\Alystar}{\ensuremath{\alpha^{\Te}_{22}}}
\newcommand{\Alzstar}{\ensuremath{\alpha^{\Te}_{33}}}
\newcommand{\Sigxxh}{\ensuremath{\left< \sigma^{\Ahbg}_{11} \right>}}
\newcommand{\Sigxyh}{\ensuremath{\left< \sigma^{\Abhg}_{12} \right>}}
\newcommand{\Sigxzh}{\ensuremath{\left< \sigma^{\Abgh}_{13} \right>}}
\newcommand{\Sigyxh}{\ensuremath{\left< \sigma^{\Ahbg}_{12} \right>}}
\newcommand{\Sigyyh}{\ensuremath{\left< \sigma^{\Abhg}_{22} \right>}}
\newcommand{\Sigyzh}{\ensuremath{\left< \sigma^{\Abgh}_{23} \right>}}
\newcommand{\Sigzxh}{\ensuremath{\left< \sigma^{\Ahbg}_{13} \right>}}
\newcommand{\Sigzyh}{\ensuremath{\left< \sigma^{\Abhg}_{23} \right>}}
\newcommand{\Sigzzh}{\ensuremath{\left< \sigma^{\Abgh}_{33} \right>}}
\newcommand{\Cxx}{C^{\Abg}_{11}}
\newcommand{\Cxxa}{C^{\Ahbg}_{11}}
\newcommand{\Cxxyy}{C^{\Abg}_{12}}
\newcommand{\Cxxyya}{C^{\Ahbg}_{12}}
\newcommand{\Cxxyyb}{C^{\Abhg}_{12}}
\newcommand{\Cxxzz}{C^{\Abg}_{13}}
\newcommand{\Cxxzza}{C^{\Ahbg}_{13}}
\newcommand{\Cxxzzg}{C^{\Abgh}_{13}}
\newcommand{\Cyy}{C^{\Abg}_{22}}
\newcommand{\Cyyb}{C^{\Abhg}_{22}}
\newcommand{\Cyyzz}{C^{\Abg}_{23}}
\newcommand{\Cyyzzb}{C^{\Abhg}_{23}}
\newcommand{\Cyyzzg}{C^{\Abgh}_{23}}
\newcommand{\Czz}{C^{\Abg}_{33}}
\newcommand{\Czzg}{C^{\Abgh}_{33}}
\newcommand{\Cyz}{C^{\Abg}_{44}}
\newcommand{\Cyzb}{C^{\Abhg}_{44}}
\newcommand{\Cyzg}{C^{\Abgh}_{44}}
\newcommand{\Cxz}{C^{\Abg}_{55}}
\newcommand{\Cxza}{C^{\Ahbg}_{55}}
\newcommand{\Cxzg}{C^{\Abgh}_{55}}
\newcommand{\Cxy}{C^{\Abg}_{66}}
\newcommand{\Cxya}{C^{\Ahbg}_{66}}
\newcommand{\Cxyb}{C^{\Abhg}_{66}}

\newcommand{\Epxx}{\left<\epsilon^{\Abg}_{11}\right>}
\newcommand{\Epyy}{\left<\epsilon^{\Abg}_{22}\right>}
\newcommand{\Epzz}{\left<\epsilon^{\Abg}_{33}\right>}

\newcommand{\Epxy}{\left<\epsilon^{\Abg}_{12}\right>}
\newcommand{\Epxz}{\left<\epsilon^{\Abg}_{13}\right>}
\newcommand{\Epyz}{\left<\epsilon^{\Abg}_{23}\right>}

\newcommand{\Ambg}{\ensuremath{(\vec{\alpha} \beta \gamma)}}
\newcommand{\Abmg}{\ensuremath{(\alpha \vec{\beta} \gamma)}}
\newcommand{\Abgm}{\ensuremath{(\alpha \beta \vec{\gamma})}}

\newcommand{\Ahbmg}{\ensuremath{(\hat{\alpha} \vec{\beta} \gamma)}}
\newcommand{\Ahbgm}{\ensuremath{(\hat{\alpha} \beta \vec{\gamma})}}
\newcommand{\Ambhg}{\ensuremath{(\vec{\alpha} \hat{\beta} \gamma)}}
\newcommand{\Ambgh}{\ensuremath{(\vec{\alpha} \beta \hat{\gamma})}}
\newcommand{\Abhgm}{\ensuremath{(\alpha \hat{\beta} \vec{\gamma})}}
\newcommand{\Abmgh}{\ensuremath{(\alpha \vec{\beta} \hat{\gamma})}}

\newcommand{\Epxxa}{\left<\epsilon^{\Ahbg}_{11}\right>}
\newcommand{\Epxxb}{\left<\epsilon^{\Abhg}_{11}\right>}
\newcommand{\Epxxbgm}{\left<\epsilon^{\Abhgm}_{11}\right>}
\newcommand{\Epxxbm}{\left<\epsilon^{\Abmg}_{11}\right>}
\newcommand{\Epxxg}{\left<\epsilon^{\Abgh}_{11}\right>}
\newcommand{\Epxxgm}{\left<\epsilon^{\Abgm}_{11}\right>}

\newcommand{\Epyya}{\left<\epsilon^{\Ahbg}_{22}\right>}
\newcommand{\Epyyam}{\left<\epsilon^{\Ambg}_{22}\right>}
\newcommand{\Epyyamg}{\left<\epsilon^{\Ambgh}_{22}\right>}
\newcommand{\Epyyb}{\left<\epsilon^{\Abhg}_{22}\right>}
\newcommand{\Epyyg}{\left<\epsilon^{\Abgh}_{22}\right>}
\newcommand{\Epyygm}{\left<\epsilon^{\Abgm}_{22}\right>}

\newcommand{\Epzza}{\left<\epsilon^{\Ahbg}_{33}\right>}
\newcommand{\Epzzabm}{\left<\epsilon^{\Ahbmg}_{33}\right>}
\newcommand{\Epzzam}{\left<\epsilon^{\Ambg}_{33}\right>}
\newcommand{\Epzzb}{\left<\epsilon^{\Abhg}_{33}\right>}
\newcommand{\Epzzbm}{\left<\epsilon^{\Abmg}_{33}\right>}
\newcommand{\Epzzg}{\left<\epsilon^{\Abgh}_{33}\right>}

\newcommand{\Epxya}{\left<\epsilon^{\Ahbg}_{12}\right>}
\newcommand{\Epxyam}{\left<\epsilon^{\Ambg}_{12}\right>}
\newcommand{\Epxyamg}{\left<\epsilon^{\Ambgh}_{12}\right>}
\newcommand{\Epxyabm}{\left<\epsilon^{\Ahbmg}_{12}\right>}
\newcommand{\Epxyb}{\left<\epsilon^{\Abhg}_{12}\right>}
\newcommand{\Epxybgm}{\left<\epsilon^{\Abhgm}_{12}\right>}
\newcommand{\Epxyamb}{\left<\epsilon^{\Ambhg}_{12}\right>}
\newcommand{\Epxybm}{\left<\epsilon^{\Abmg}_{12}\right>}
\newcommand{\Epxyg}{\left<\epsilon^{\Abgh}_{12}\right>}
\newcommand{\Epxygm}{\left<\epsilon^{\Abgm}_{12}\right>}

\newcommand{\Epxza}{\left<\epsilon^{\Ahbg}_{13}\right>}
\newcommand{\Epxzabm}{\left<\epsilon^{\Ahbmg}_{13}\right>}
\newcommand{\Epxzam}{\left<\epsilon^{\Ambg}_{13}\right>}
\newcommand{\Epxzamg}{\left<\epsilon^{\Ambgh}_{13}\right>}
\newcommand{\Epxzagm}{\left<\epsilon^{\Ahbgm}_{13}\right>}
\newcommand{\Epxzb}{\left<\epsilon^{\Abhg}_{13}\right>}
\newcommand{\Epxzbgm}{\left<\epsilon^{\Abhgm}_{13}\right>}
\newcommand{\Epxzbm}{\left<\epsilon^{\Abmg}_{13}\right>}
\newcommand{\Epxzg}{\left<\epsilon^{\Abgh}_{13}\right>}
\newcommand{\Epxzgm}{\left<\epsilon^{\Abgm}_{13}\right>}

\newcommand{\Epyza}{\left<\epsilon^{\Ahbg}_{23}\right>}
\newcommand{\Epyzagm}{\left<\epsilon^{\Ahbgm}_{23}\right>}
\newcommand{\Epyzam}{\left<\epsilon^{\Ambg}_{23}\right>}
\newcommand{\Epyzamg}{\left<\epsilon^{\Ambgh}_{23}\right>}
\newcommand{\Epyzb}{\left<\epsilon^{\Abhg}_{23}\right>}
\newcommand{\Epyzbm}{\left<\epsilon^{\Abmg}_{23}\right>}
\newcommand{\Epyzbgm}{\left<\epsilon^{\Abhgm}_{23}\right>}
\newcommand{\Epyzg}{\left<\epsilon^{\Abgh}_{23}\right>}
\newcommand{\Epyzbmg}{\left<\epsilon^{\Abmgh}_{23}\right>}
\newcommand{\Epyzgm}{\left<\epsilon^{\Abgm}_{23}\right>}

  \section{Introduction}
  Exact relations for the effective elastic properties of two-component composites can be classified into three types.  The first type consists of relations that have been determined from the similarity of the two-dimensional stress and strain fields for certain types of materials.  These exact relations are called duality relations~\cite{Helsing97}.  The second type of exact relations, called translation-based relations, state that if a constant quantity is added to the elastic moduli of the component materials then the effective elastic moduli are also ``translated'' by the same amount.  Microstructure independent exact relations, valid for special combinations of the elastic properties of the components, form the third category~\cite{Milton97}.  The known exact relations are directly applicable only to a limited range of properties of the components.  Therefore the utility of these relations lies not only in determining the effective elastic properties of a small range of composites but also in evaluating the accuracy of numerical and analytical methods of computing effective properties.  In this work, predictions from exact relations are compared with estimates from finite element calculations, the generalized method of cells (GMC)~\cite{Aboudi96_1}, and the recursive cell method (RCM)~\cite{Banerjee02c}.  The goal is to assess the effectiveness of these relations in evaluating the accuracy of the three numerical methods, especially with regard to high modulus contrast materials such as polymer bonded explosives.

  Five exact relations are explored in this work.  The first is a duality-based identity for the effective shear modulus that is valid for phase-interchangeable materials~\cite{Milton02}.  The second is a set of duality relations that are valid for materials that are rigid with respect to shear~\cite{Helsing97}.  Two translation-based relations are explored next - the CLM theorem~\cite{Cherk92} and a relation for symmetric composites with equal bulk modulus~\cite{Milton02}.  The microstructure independent Hill's relation~\cite{Hill64} is explored last.

    \section{Phase interchange identity}
\begin{figure}[t]
     \begin{center}
	\scalebox{0.45}{\includegraphics{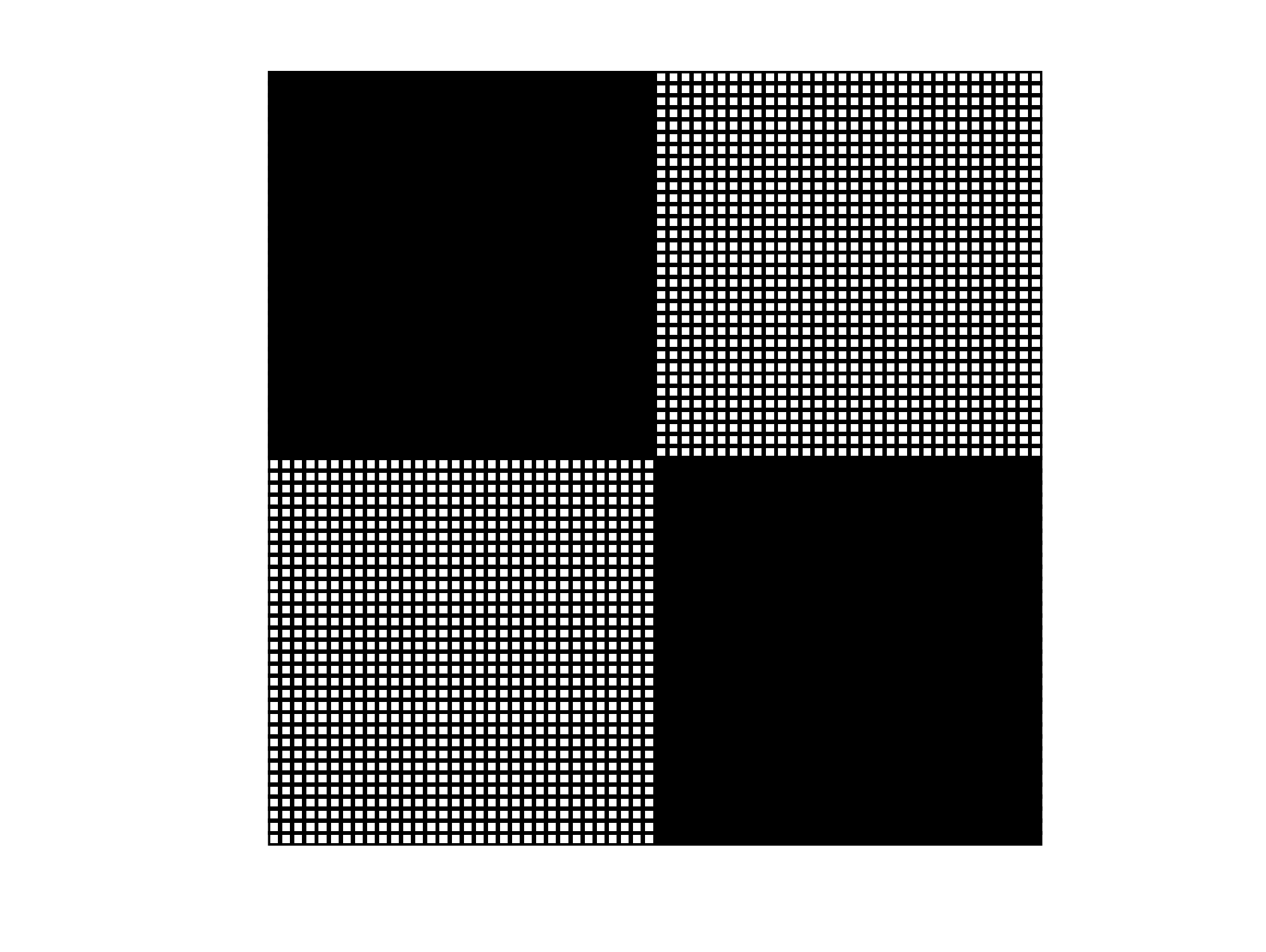}}
     \end{center}
 	\caption{Representative volume element for a checkerboard composite.}
        \label{fig:checker_64}
  \end{figure}

  A symmetric composite is one that is invariant with respect to interchange of the components.  A checkerboard, as shown in Figure~\ref{fig:checker_64}, is an example of a symmetric composite.
  The phase interchange identity~\cite{Milton02} for the effective shear modulus of a symmetric two-dimensional two-component isotropic composite is a duality-based exact relation that states that
  \begin{equation}
      G_{\Te} = \sqrt{G_1 G_2}
  \end{equation}
  where $G_1$, $G_2$ are the shear moduli of the two components and $G_{\Te}$ is the effective shear modulus.

  The phase interchange identity is valid only for isotropic composites.  In a finite-sized representative volume element (RVE) for a checkerboard composite the shear modulus is not the same all directions and hence isotropy is not achieved.  The two-dimensional stress-strain relation for such a RVE with ``square symmetry'' can be written as
  \begin{equation}
    \label{eq:squareSymmExact}
    \left[\begin{array}{c} \Sigx \\ \Sigy \\ \Tauxy \end{array}\right]  =
    \left[\begin{array}{lll} K + \mu^1 & K - \mu^1 & 0 \\ K - \mu^1 & K + \mu^1 & 0 \\ 
		0 & 0 & \mu^2 \end{array}\right]
    \left[\begin{array}{c} \Epsx \\ \Epsy \\ 2\Epsxy \end{array} \right]
  \end{equation}
  where $\Sigx$, $\Sigy$, $\Tauxy$ are the stresses; $\Epsx$, $\Epsy$, $\Gamxy$ are the strains; $K$ is the two-dimensional bulk modulus, $\mu^1$ is the shear modulus when shear is applied along the diagonals of the RVE, and $\mu^2$ is the shear modulus for shear along the edges of the RVE. 

  The numerical verification of the phase interchange identity, therefore, requires that the components of the composite be chosen so that the difference between $\mu^1$ and $\mu^2$ for the composite is minimal.  This implies that the components should have a weak modulus contrast.

  Numerical estimates of the effective elastic properties of the checkerboard composite shown in Figure~\ref{fig:checker_64} were obtained using finite elements (FEM), the recursive cell method (RCM) and the generalized method of cells (GMC).  Following the requirement of low modulus contrast, both components were assigned a Young's modulus of 15,300 MPa.  The Poisson's ratio of the first component was fixed at 0.32 while that of the second component was varied from 0.1 to 0.49.  The FEM calculations were performed using a mesh of 256$\times$256 four-noded square elements.  The RCM calculations used a grid of 64$\times$64 subcells with blocks of 2$\times$2 subcells and each subcell was modeled using one nine-noded element.  The GMC calculations used 64$\times$64 square subcells to discretize the RVE.

\begin{figure}[t]
     \begin{center}
	\scalebox{0.65}{\includegraphics{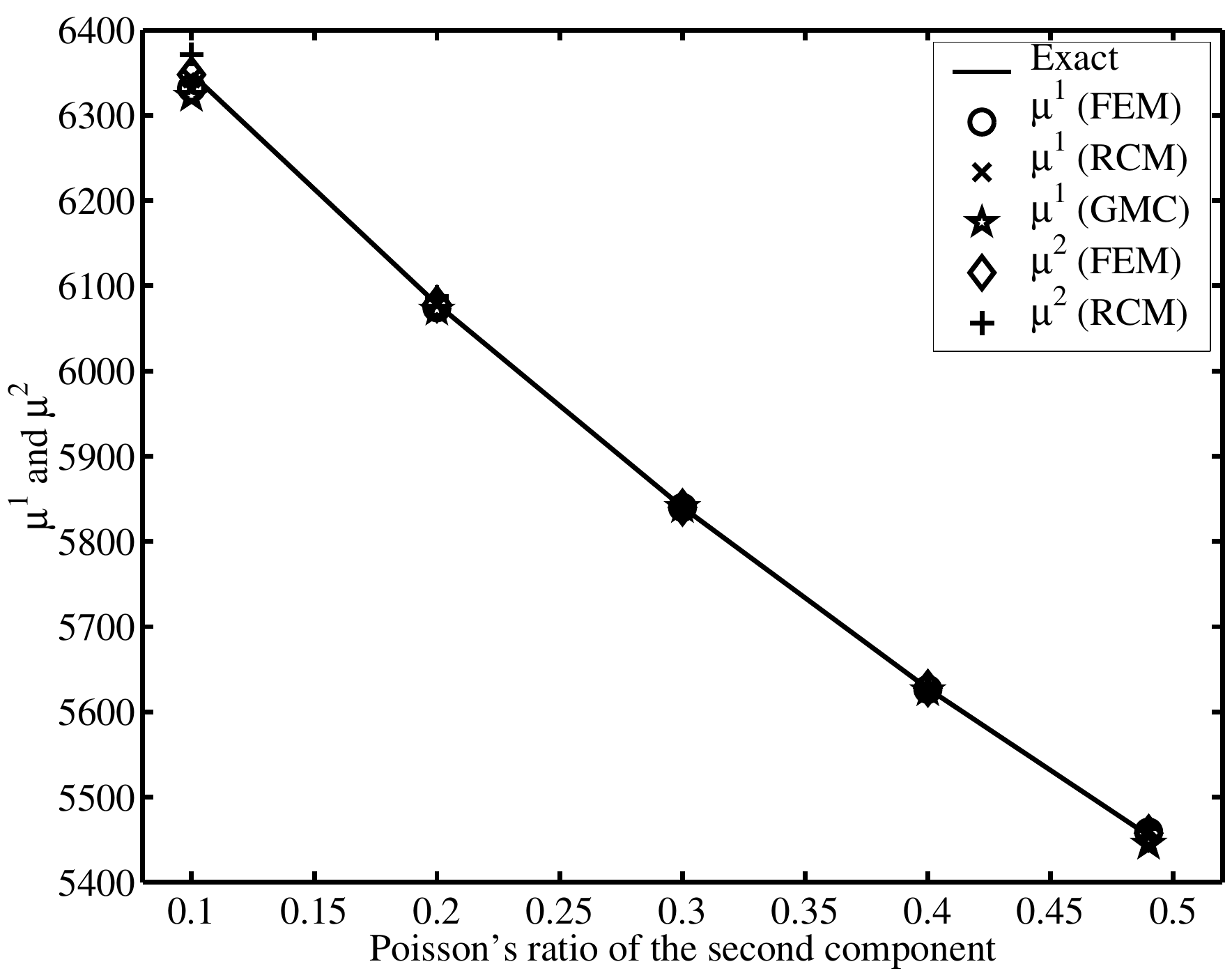}}
     \end{center}
 	\caption[Validation of FEM, RCM and GMC using the phase interchange 
		identity for a checkerboard composite.]
 	        {Validation of FEM, RCM and GMC using the phase interchange 
		identity for a checkerboard composite.}
        \label{fig:checkerIsotropic}
  \end{figure}

  Figure~\ref{fig:checkerIsotropic} shows a comparison of the exact effective shear modulus for the checkerboard composite with estimates of $\mu^1$ and $\mu^2$ from the three numerical approaches.
  The results show that all the three methods perform well (the maximum error is 0.1\%) in predicting the effective shear modulus when the modulus contrast is small, i.e., when the composite is nearly isotropic.  It can also be observed that the values of $\mu^1$ and $\mu^2$ are within 1\% of each other for the chosen component moduli.

  \subsection{Range of applicability}
  The question that arises at this point is whether the three numerical approaches can predict the phase interchange identity for larger modulus contrasts.  Numerical calculations have been performed on the checkerboard microstructure to explore this issue.  The first component of the checkerboard was assigned a Young's modulus of 15,300 MPa and a Poisson's ratio of 0.32.  For the second component, the Poisson's ratio was fixed at 0.49 and the Young's modulus was varied from 0.7 MPa to 7000 MPa.

\begin{figure}[t]
     \begin{center}
	\scalebox{0.65}{\includegraphics{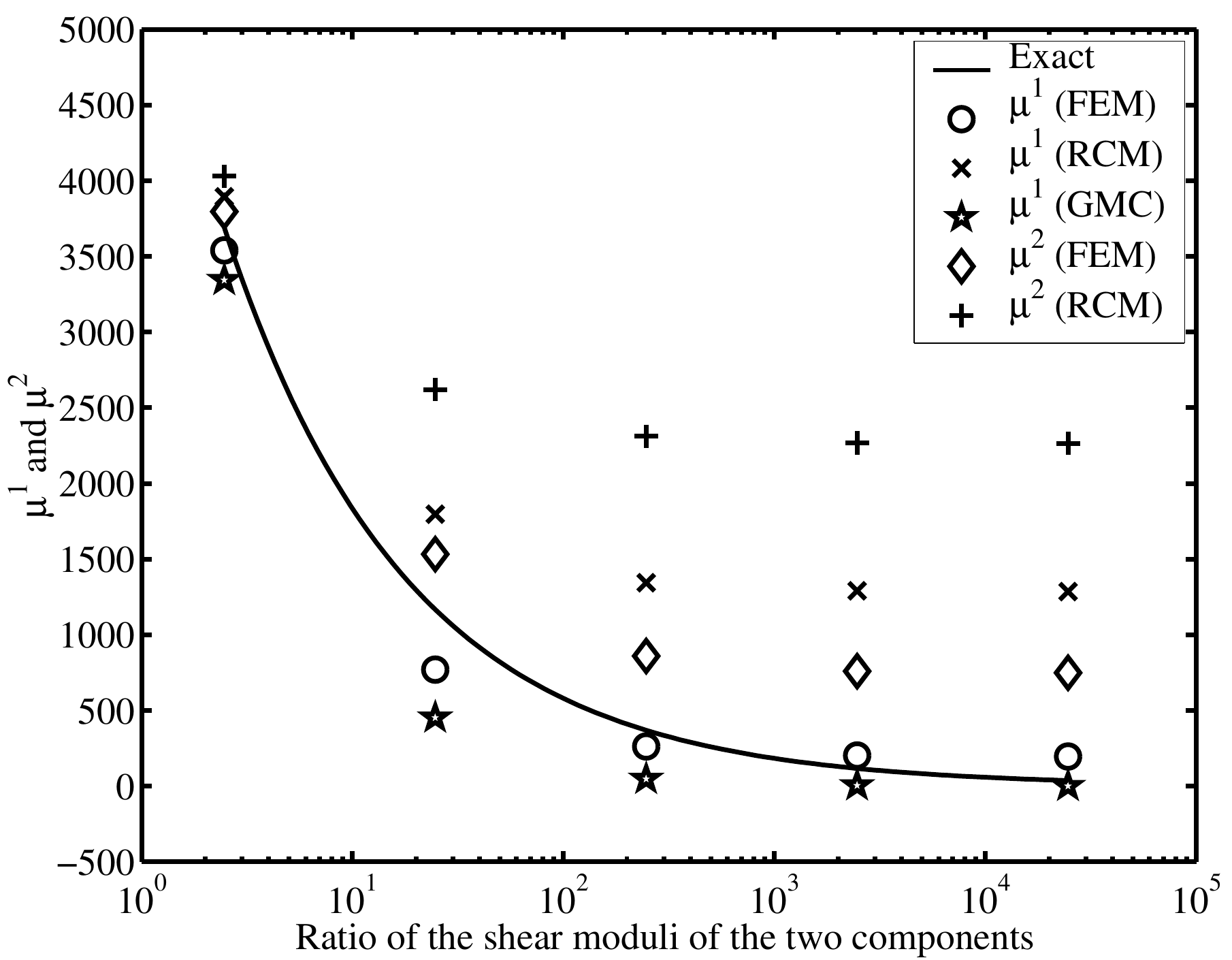}}
     \end{center}
 	\caption[Variation of effective shear moduli with modulus contrast
		for a checkerboard composite.]
 	        {Variation of effective shear moduli with modulus contrast
		for a checkerboard composite.}
        \label{fig:checkerContrast}
  \end{figure}

  Figure~\ref{fig:checkerContrast} shows plots of the effective $\mu^1$ and $\mu^2$ versus shear modulus contrast for a checkerboard RVE.
  The plots confirm that when the modulus contrast between the components of the checkerboard exceeds 2, the material can no longer be considered isotropic since the values of $\mu^1$ and $\mu^2$ are considerably different from each other.  However, the values of $\mu^1$ predicted by FEM are quite close to the effective shear modulus $G_{\Te}$ predicted by the phase interchange identity.  This result suggests that the simulation of a diagonal shear may not be necessary to predict the effective shear modulus of an isotropic composite when the finite element approach is used.  It also implies that the phase interchange identity can be used for a much larger range of modulus contrasts.  The effective shear moduli predicted by GMC are considerably lower than that from the exact relation while the values from RCM are consistently higher.  The RCM estimates worsen with increasing modulus contrast.  If only the value of $\mu^1$ is examined, the phase interchange identity indicates that the FEM approach is much more accurate than the GMC and RCM approaches.  However, it is difficult to choose between GMC and RCM for high modulus contrasts composites.  While the exact value of $\mu^1$ is 10 times the value predicted by GMC, the corresponding RCM estimate is 10 times the exact value.  These results confirm the findings of detailed numerical studies on high modulus contrast, high volume fraction polymer bonded explosives~\cite{Banerjee02a,Banerjee02b,Banerjee02c}.

  \subsection{Convergence of FEM calculations}
\begin{figure}[t]
     \begin{center}
	\scalebox{0.63}{\includegraphics{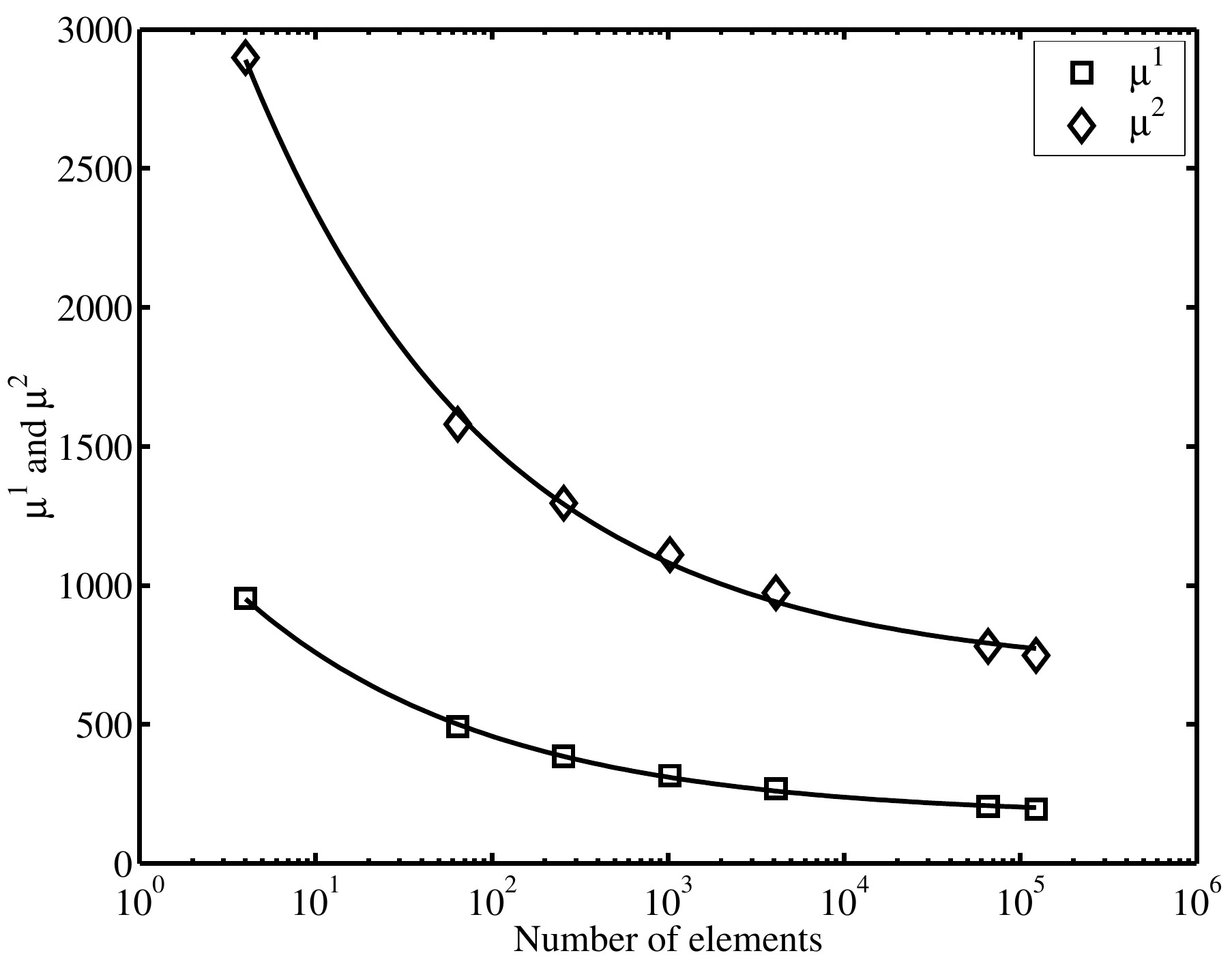}}
 	\caption[Convergence of effective moduli predicted by FEM
		 analyses with increase in mesh refinement for a checkerboard 
		 composite with shear modulus contrast of 25,000.]
 	        {Convergence of effective moduli predicted by FEM with increase 
                 in mesh refinement for a checkerboard 
		 composite with a shear modulus contrast of 25,000.}
        \label{fig:checkerFEconv}
     \end{center}
  \end{figure}

  The checkerboard material provides an extreme case to test the convergence of the FEM solution because the corner singularities lead to high stresses that can only be resolved with refined meshes.  Figure~\ref{fig:checkerFEconv} shows the convergence of the effective $\mu^1$ and $\mu^2$ with increasing mesh refinement for a checkerboard with a shear modulus contrast of about 25,000.
  The effective $\mu^1$ converges to a steady value when 128$\times$128 elements are used to discretize the RVE.  The shear modulus $\mu^2$ reaches a steady value when 256$\times$256 elements are used.  This is why the finite element calculations in this work were performed using 256$\times$256 elements or more.

  RCM uses a finite element approach to homogenize blocks of subcells.  When blocks of 2$\times$2 subcells are used, some of these blocks can resemble checkerboards - especially at the first level of recursion for a two-component composite.  The finite element convergence result suggests that RCM may overestimate the effective shear moduli by a factor of two if a block of four subcells is simulated using only four finite elements.

  \section{Materials rigid in shear}
  The stress-strain response of two-dimensional materials that are rigid with respect to shear can be represented by
  \begin{equation}
     \left[\begin{array}{c} \Epsx \\ \Epsy \\ \Gamxy \end{array}\right] = 
     \left[\begin{array}{ccc}
        S_{11} & S_{12} & 0 \\ S_{12} & S_{22} & 0 \\ 0 & 0 & 0 
     \end{array}\right]
     \left[\begin{array}{c} \Sigx \\ \Sigy \\ \Tauxy \end{array}\right]
  \end{equation}
  where $\Sigx$, $\Sigy$, $\Tauxy$ are the stresses; $\Epsx$, $\Epsy$ and $\Gamxy$ are the strains, and $S_{ij}$ are the components of the compliance matrix.

  Two duality-based exact relations that are valid for two-component composites composed of such materials are~\cite{Helsing97}: 
  \begin{description}
     \item[{Relation RS1}] {
	If $S_{11} S_{22} - (S_{12})^2 = \Delta$ for each phase (where $\Delta$
        is a constant), then the effective compliance tensor also satisfies 
        the same relationship, i.e., $\Seff{11} \Seff{22} - (\Seff{12})^2 = \Delta_{\Te}$.
	This relation is true for all microstructures.
     }
     \item[{Relation RS2}] {
	If the compliance tensors of the two phases are of the form
        $\mathbf{S}_1 = \alpha_1 \mathbf{A}$ and 
        $\mathbf{S}_2 = \alpha_2 \mathbf{A}$ where $\mathbf{A}$ is a constant
        matrix, then the effective compliance tensor of a checkerboard of the
        two phases satisfies the relation 
        $ \det{\mathbf{S}_{\Te}} = \Seff{11} \Seff{22} - (\Seff{12})^2 = 
		\alpha_1\alpha_2(A_{11}A_{22} - (A_{12})^2)$.
     }
  \end{description}
  
  \subsection{Relation RS1}
\begin{figure}[b]
     \begin{center}
	\scalebox{0.5}{\includegraphics{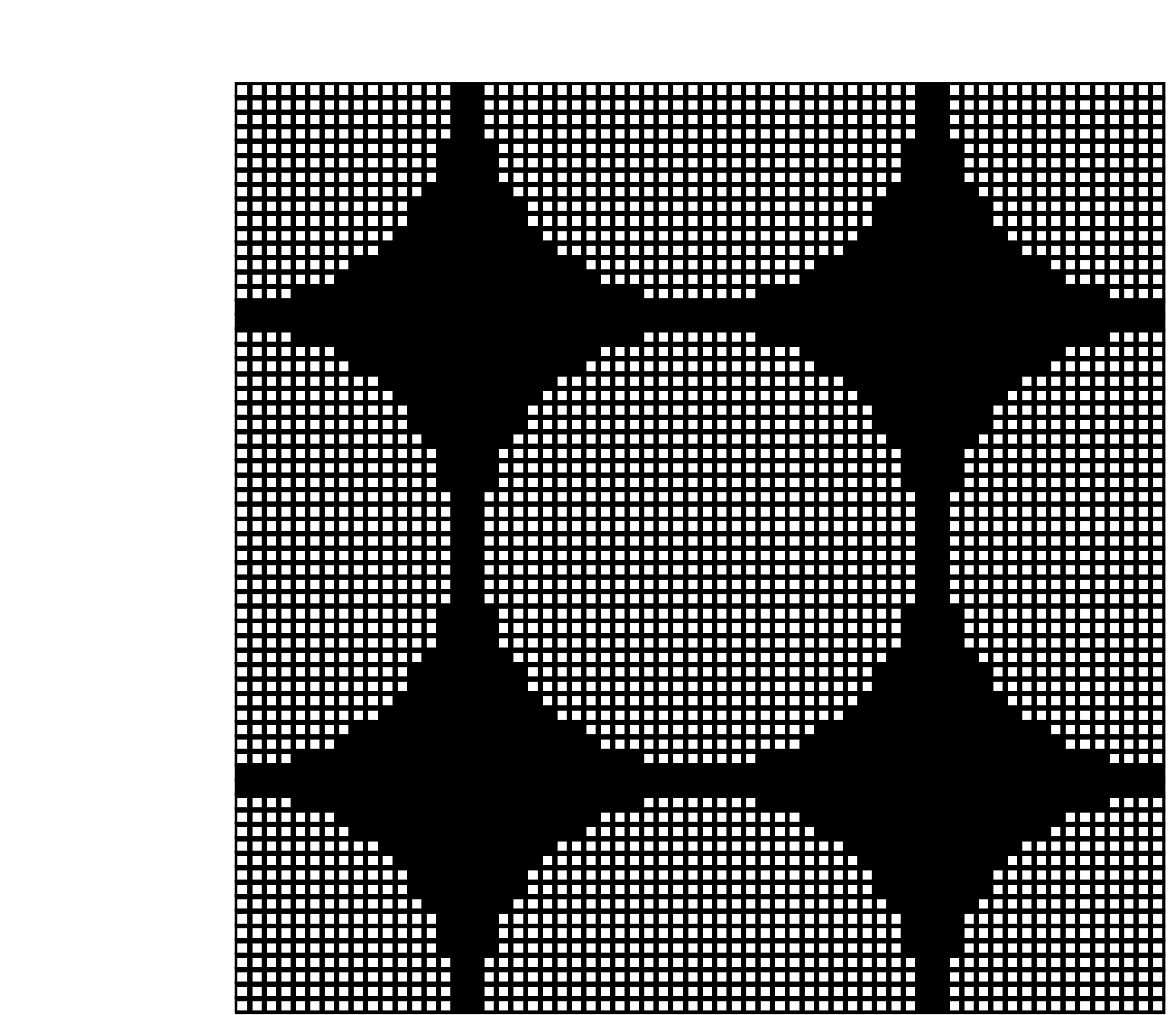}}
     \end{center}
 	\caption{RVE for a square array of disks.}
        \label{fig:diskArr70_64}
  \end{figure}

  Figure~\ref{fig:diskArr70_64} shows a square array of disks occupying an area fraction of 0.7.
  Numerical experiments have been performed on this array of disks to check if Relation RS1 can be reproduced by finite element analyses, GMC and RCM.  The $\mathbf{S}$ matrices that have been used for the disks (superscript $1$) and the matrix (superscript $2$), and the corresponding values of $\Delta$ are shown below.  These matrices have been chosen so that the value of $\Delta$ is constant.  
  \begin{equation*}  
     \mathbf{S}_1  = \left[\begin{array}{ccc} 1000 & -300 & 0 \\
	-300 & 1000 & 0 \\ 0 & 0 & 0.001 \end{array}\right],~  
	\Delta = 9.1\times10^5,
  \end{equation*}  
  and
  \begin{equation*}  
     \mathbf{S}_2  = \left[\begin{array}{ccc} 1094.3 & -536.21 & 0 \\
	-536.21 & 1094.3 & 0 \\ 0 & 0 & 0.001 \end{array}\right],~ 
	\Delta = 9.1\times10^5.
  \end{equation*}  
  The shear modulus for both materials is $1000$ (arbitrary units) - around $10^6$ times the Young's modulus.  Higher values of shear modulus have been tested and found not to affect the effective stiffness matrix significantly.

  Table~\ref{tab:diskArr70_64} shows the values of $\Seff{11}$, $\Seff{12}$ and $\Delta_{\Te}$ calculated using finite elements (350$\times$350 elements), GMC (64$\times$64 subcells) and RCM (256$\times$256 subcells).  
  \begin{table}[b]
	\caption[Two-dimensional effective compliance matrix for a square array of disks.] 
	        {Two-dimensional effective compliance matrix for a square array of disks.} \medskip
        \label{tab:diskArr70_64}
     \begin{center}
	\begin{tabular}{lcccc}
	   \hline
           & $\Seff{11}$ & $\Seff{12}$ & $\Delta_{\Te}$ \Ti{5} & $\Delta_{\Te}$/$\Delta$ \\
	   \hline
           FEM & 850.35 &  -536.32 &   4.35 &   0.48 \\
           RCM & 847.25 &  -538.17 &   4.28 &   0.47 \\
           GMC & 871.75 &  -517.14 &   4.93 &   0.54 \\
	   \hline
        \end{tabular}
     \end{center}
  \end{table}
  The ratio of the calculated $\Delta_{\Te}$ to the original $\Delta$ are also shown in the table.  The modulus contrast between the two components of the composite is small, so the calculated effective properties are expected to be accurate (based on the results on the phase interchange identity for shear moduli).  However, the results in Table~\ref{tab:diskArr70_64} show that all the three numerical methods predict values of $\Delta_{\Te}$ that are around half the original $\Delta$.  These results imply that all three methods (FEM, GMC and RCM) overestimate the effective normal stiffness of the array of disks.  Relation RS1 for materials rigid in shear may therefore be a very sensitive test of the accuracy of numerical methods even though the modulus contrast that can be used is small.

  \subsection{Relation RS2}
  The second duality relation for materials that are rigid in shear requires (Relation RS2) is valid for the checkerboard geometry shown in Figure~\ref{fig:checker_64}. The following values of the elastic properties have been used to test the accuracy of FEM, RCM and GMC in predicting this relation.
  \begin{eqnarray*}
     \mathbf{S}_1 & = 100 \left[\begin{array}{cc} 10 & -3 \\ -3 & 10 \end{array}\right] ~;~
     \mathbf{S}_2 = 1000 \left[\begin{array}{cc} 10 & -3 \\ -3 & 10 \end{array}\right] \\
     \alpha_1 & = 100 ~;~ \alpha_2 = 1000 \\ 
     \mathbf{A} & = \left[\begin{array}{cc} 10 & -3 \\ -3 & 10 \end{array}\right] 
  \end{eqnarray*}
  The duality relation requires that the effective compliance matrix of the checkerboard composite should be such that
  \begin{equation*}
     \det(\mathbf{S}_{\Te}) = \Seff{11} \Seff{22} - (\Seff{12})^2 = 9.10\times10^6~.
  \end{equation*}

  The FEM calculations were performed using 350$\times$350 four-noded elements, the RCM calculations used 64$\times$64 subcells (blocks of 2$\times$2 subcells) and the GMC calculations used 64$\times$64 subcells too.  The results from these three methods are tabulated in Table~\ref{tab:checker_64}.  
  \begin{table}[t]
     \begin{center}
	\caption{Effective compliance matrix for a checkerboard composite with 
                 components rigid in shear.}\medskip
        \label{tab:checker_64}
	\begin{tabular}{lcccc}
	   \hline
           & $\Seff{11}$ & $\Seff{12}$ & $\det(\mathbf{S}_{\Te}) \Ti{6}$ & 
             $\det(\mathbf{S}^{\Te})/\alpha_1\alpha_2\det{\mathbf{A}}$  \\
	   \hline
           FEM & 3282 & -2004 & 6.75 & 0.74 \\
           RCM & 1655 & -7090 & 2.23 & 0.24 \\
           GMC & 5007 & -2146 & 2.05 & 2.25 \\
	   \hline
        \end{tabular}
     \end{center}
  \end{table}
  The finite element calculations lead to quite an accurate effective compliance matrix and the deviation from the exact result is only around 25\%.  The GMC calculations overestimate the compliance matrix and the determinant of the compliance matrix is around 2.3 times higher than the exact result.  On the other hand, the RCM calculations predict a compliance matrix that has a determinant that is only around 20% of the exact value.

  \section{The CLM theorem}
  The Cherkaev, Lurie and Milton (CLM) theorem is a well known ``translation'' based exact relation for two-component planar composites~\cite{Cherk92}.  For a two-dimensional two-component isotropic composite, this theorem can be stated as follows.  

  Let the isotropic bulk moduli of the components be $K_1$ and $K_2$.  Let the shear moduli of the two components be $G_1$ and $G_2$.  The effective bulk and shear modulus of a two-dimensional composite made of these two components are $K_{\Te}$ and $G_{\Te}$ respectively.  Let us now create two new materials that are ``translated'' from the original component materials by a constant amount $\lambda$.  That is, let the bulk and shear moduli of the translated component materials be given by
  \begin{eqnarray*}
    1/K^T_1 = 1/K_1 - \lambda~;~ 1/K^T_2 = 1/K_2 - \lambda~; \\
    1/G^T_1 = 1/G_1 + \lambda~;~ 1/G^T_2 = 1/G_2 + \lambda~.
  \end{eqnarray*}
  The CLM theorem states that the effective bulk and shear moduli of a two-dimensional composite of the two translated materials, having the same microstructure as the original composite, are given by
  \begin{equation}
     1/K^T_{\Te} = 1/K_{\Te} - \lambda~;~
     1/G^T_{\Te} = 1/G_{\Te} + \lambda. \label{eq:CLMIsotropic}
  \end{equation}

  The requirement of isotropy can be satisfied approximately by choosing component material properties that are close to each other.  Since our goal is to determine how well GMC and RCM perform for high modulus contrast, choosing materials with small modulus contrast is not adequate.  Another alternative is to choose a RVE that represents a hexagonal packing of disks.  However, such an RVE is necessarily rectangular and cannot be modeled using RCM in its current form. It should be noted that RCM can easily be modified to deal with elements that are not square and hence to model rectangular regions.

  Another problem in the application of the CLM theorem is that the value of $\lambda$ has to be small if the difference between the original and the translated moduli is large and vice versa.  If the value of $\lambda$ is small, floating point errors can accumulate and exceed the value of $\lambda$.  On the other hand, if $\lambda$ is large, the original and the translated moduli are very close to each other and the difference between the two can be lost because of errors in precision.  Hence, the numbers have to be chosen carefully keeping in mind the limits on the value of the Poisson's ratio.
  
The translation relation has been tested on the square array of disks occupying a volume fraction of 0.70 from Figure~\ref{fig:diskArr70_64}.  This RVE exhibits square symmetry, i.e., the shear moduli $\mu^1$ and $\mu^2$ shown in equation~(\ref{eq:squareSymmExact}) are not equal.  A unique value of the effective shear modulus cannot be calculated for this RVE.  Instead, he value of the effective translated shear modulus is calculated from equation~(\ref{eq:CLMIsotropic}) by first setting $G_{\Te}$ equal to $\mu^1$ and then to $\mu^2$.  These ``exact'' values are compared with the $\mu^1$ and $\mu^2$ values predicted using finite element analyses, GMC and RCM.

  The original set of elastic moduli for the RVE is chosen to reflect the elastic moduli of the constituents of polymer bonded explosives.  These moduli are then translated by a constant $\lambda = 0.001$.  The original and the translated constituent two-dimensional moduli are shown in Table~\ref{tab:clmModuli} (phase 'p' represents the particles and phase 'b' represents the binder).  It can be observed that the translation process creates quite a large change in the bulk modulus of the particles.
  \begin{table}[b]
    \begin{center}
      \caption[Original and translated two-dimensional constituent moduli 
	       for checking the CLM condition.]
              {Original and translated two-dimensional constituent moduli 
		 for checking the CLM condition.} \medskip
      \label{tab:clmModuli}
      \begin{tabular}{lcccccc}
	\hline
	& $K_p$ & $G_p$ & $K_b$ & $G_b$ & $K_p/K_b$ & $G_p/G_b$  \\
	&\Ti{2}&\Ti{2}& & & \Ti{2}&\Ti{2}\\
	\hline
	Original  & 9.60 & 4.80 & 10.07 & 0.20 &
		0.95 & 23.8 \\
        Translated & 240.0 & 3.24 & 10.17 & 0.20  &
		23.5 & 16.1 \\
	\hline
      \end{tabular}
    \end{center}
  \end{table}
  Table~\ref{tab:clmcheck} shows the effective bulk and shear moduli of the original and the translated material calculated using finite elements (350$\times$350 elements), GMC (64$\times$64 subcells) and RCM (256$\times$256 subcells).  The values of $\lambda_{\text{err}}$ shown in the table have been calculated using the equation
  \begin{eqnarray*}
    \lambda_{\text{err}} & = (\lambda/0.001-1)\times 100, \\
    \lambda = 1/K_{\Te} - 1/K^T_{\Te} = 1/\mu^{i(T)}_{\Te} - 1/\mu^{i}_{\Te}.
  \end{eqnarray*}
  \begin{table}[b]
    \begin{center}
      \caption{Comparison of effective moduli for the original and the
		translated composites.} \medskip
      \label{tab:clmcheck}
      \begin{tabular}{lccccccccc}
	\hline
        & \multicolumn{3}{c}{$K_{\Te}$} & \multicolumn{3}{c}{$\mu^1_{\Te}$}
          & \multicolumn{3}{c}{$\mu^2_{\Te}$} \\
        & Orig. & Trans. & $\lambda_{\text{err}}$(\%) & 
	  Orig. & Trans. & $\lambda_{\text{err}}$(\%) & 
	  Orig. & Trans. & $\lambda_{\text{err}}$(\%) \\
	\hline
        FEM & 36.4 & 37.8 & -0.8 & 10.1 & 10 & -3.1 & 0.9 & 0.9 & 22\\
        RCM & 42.5 & 44.5 & 6.1 & 29.8 & 29 & -6.9 & 1.3 & 1.3 & -292 \\ 
        GMC & 34.0 & 35.1 & -1.3 & 3.8 & 3.8 & 5.3 & 0.7 & 0.7 & 30 \\
	\hline
      \end{tabular}
    \end{center}
  \end{table}

  Even though the modulus contrast between the two components of the composite is high, the effective properties predicted by FEM, GMC and RCM are close to each other in magnitude.  The effective moduli of the translated composite are also quite close to that of the original composite as predicted by the CLM condition.  The interesting fact is that all the three methods satisfy the CLM condition and the error is small (as seen by the values of $\lambda_{\text{err}}$.  Of the three methods, FEM and GMC produce the least error while RCM produces the most error.

  \section{Composites with equal bulk modulus}
  The translation procedure can also be used to generate an exact solution for the effective shear modulus of two-dimensional symmetric two-component composites with both components having the same bulk modulus~\cite{Milton02}.  This relation is
  \begin{eqnarray}
     K_{\Te} & =~K~=~K_1~=~K_2 \nonumber \\
     G_{\Te} & = \frac{K}{-1 + \sqrt{\left(1+\frac{K}{G_1}\right).
	\left(1+\frac{K}{G_2}\right)}}
     \label{eq:sameBulk}
  \end{eqnarray}
  This relation has been tested on the checkerboard model shown in Figure~\ref{fig:checker_64} using the component material properties given in Table~\ref{tab:sameBulk}.  The exact effective properties for the composite, calculated using equation~(\ref{eq:sameBulk}), are also given in the table.  The values of the effective moduli calculated using finite elements (FEM), GMC and RCM are also shown in Table~\ref{tab:sameBulk}.  
  \begin{table}[b]
    \begin{center}
      \caption[Component properties, exact effective properties and numerically computed effective 
               properties for two-component symmetric composite with equal component bulk moduli.]
              {Component properties, exact effective properties and
		numerically computed effective properties
		for two-component symmetric composite 
		with equal component bulk moduli.}\medskip
      \label{tab:sameBulk}
      \begin{tabular}{lcccc}
	\hline
	& $E$ & $\nu$ & $K$ & $G$ \\
	&\Ti{2}&  & \Ti{3} & \Ti{2} \\
	\hline
	Component 1   & 25.00 & 0.25 & 2.0 & 10.0 \\
	Component 2   & 1.19 & 0.49 & 2.0 & 0.4 \\
	Composite     & 5.12 & 0.46 & 2.0 & 1.76 \\
	\hline
	\multicolumn{4}{c}{}\\
      \end{tabular}
      \begin{tabular}{lccccc}
	\hline
	& $K_{\Te}$ & $\mu^1_{\Te}$ & Diff. & $\mu^2_{\Te}$ & Diff.\\
	& \Ti{2}& \Ti{2} & \% & \Ti{2} & \% \\
	\hline
	FEM & 20 & 1.29 & -26.8 & 2.54 & 44.4 \\
        GMC & 20 & 0.77 & -56.3 & 0.77 & -56.3 \\
        RCM & 20 & 2.96 & 68.0 & 4.41 & 150.9 \\
	\hline
      \end{tabular}
    \end{center}
  \end{table}

  These results show that the effective two-dimensional bulk modulus is calculated correctly by all the three methods.  However, the shear moduli calculated for the checkerboard microstructure are quite different from the exact result.  This exact result also shows that the FEM calculations are the most accurate, followed by GMC and then RCM.  The values of $\mu^1_{\Te}$ are also found to most closely approximate the value of $G_{\Te}$.

  \section{Hill's equation}
  Hill's equation~\cite{Hill64} is an exact relation that is independent of microstructure.  This equation is valid for composites composed of isotropic components that have the same shear modulus.  For a two-dimensional two-component composite, this equation can be written as
  \begin{equation}
    \frac{1}{K_{\Te}+G} = \frac{f_p}{K_p+G} + \frac{f_b}{K_b+G}
  \end{equation}
  where $f$ represents a volume fraction, $K$ represents a bulk modulus, and $G$ represents a shear modulus.  The subscript '$p$' represents a particle property, '$b$' represents a binder property, and '$\Te$' represents the effective property of the composite.

  This relationship is verified using the RVE containing an array of disks occupying 70\% of the volume that is shown in Figure~\ref{fig:diskArr70_64}.  Table~\ref{tab:hillrel} shows the properties of the two components used to compare the predictions of finite elements, GMC and RCM with the exact value of bulk modulus predicted by Hill's equation. It should be noted that the materials chosen are not quite representative of polymer bonded explosive materials.
  \begin{table}[b]
    \begin{center}
      \caption{Phase properties used for testing Hill's equation
	       and the exact effective moduli of the composite.}\medskip
      \label{tab:hillrel}
      \begin{tabular}{lccccc}
	\hline
	 & Vol. & $E$ & $\nu$ & $G$ & $K$  \\
	 & Frac. & \Ti{3} & & \Ti{3} & \Ti{3} \\
	\hline
	Disks  & 0.7 & 3.00 & 0.25 & 1.20 & 2.40 \\
	Binder & 0.3 & 3.58 & 0.49 & 1.20 & 60.00 \\
        Composite & 1.0 & 3.22 & 0.34 & 1.20 & 3.82 \\
	\hline
      \end{tabular}
    \end{center}
  \end{table}

  Since the modulus contrast is small, the square array of disks is expected to exhibit nearly isotropic behavior.  Therefore, the predictions of finite elements, GMC and RCM are expected to be close to the exact values of the effective properties of the composite.  The numerically calculated values of the effective two-dimensional bulk and shear moduli of the composite are shown in Table~\ref{tab:hillrelNum}.  The percentage difference of the effective bulk modulus from the exact value is also shown in the table.  
  \begin{table}[t]
    \begin{center}
      \caption[Numerically computed effective properties for a square array of disks with 
               equal component shear moduli.]
              {Numerically computed effective properties for a square 
	       array of disks with equal component shear moduli.}\medskip
      \label{tab:hillrelNum}
      \begin{tabular}{l|cc|cc}
	\hline
	& $K_{\Te}$ & \% Diff. & $\mu^1_{\Te}$ & $\mu^2_{\Te}$ \\
	& ($\times 10^3$) & & ($\times 10^3$) & ($\times 10^3$) \\
	\hline
	FEM & 3.98 & 4.4 & 1.20 & 1.20 \\
        RCM & 3.92 & 2.7 & 1.20 & 1.20 \\
        GMC & 3.66 & -4.2 & 1.20 & 1.20 \\
	\hline
      \end{tabular}
    \end{center}
  \end{table}

  The effective shear moduli predicted by all the three methods are exact.  In case of the effective bulk moduli, the RCM predictions are the most accurate followed by GMC and the finite element based calculations.  The finite element based calculations overestimate the effective two-dimensional bulk modulus by around 4.4\% while GMC underestimates the bulk modulus by around 4.2\%.  Since the error of estimation of all the three methods is small, it is suggested that all three methods are accurate for low contrasts in the shear modulus.  However, Hill's equation does not appear to be suitable for determining the best numerical method of the three.

    \section{Summary and conclusions}
  Predictions from the phase interchange identity for the shear modulus are closely approximated by the finite element approach (FEM), the recursive cell method (RCM) and the generalized method of cells (GMC) for checkerboard composites with low modulus contrast.  However, for higher modulus contrasts the FEM approximations of shear moduli are the most accurate.  The RCM predictions overestimate the shear modulus while GMC underestimates the shear modulus.  The exact relations for materials that are rigid in shear show that all three numerical techniques are inaccurate.  The exact relation for this class of materials that is applicable to checkerboard materials shows that the FEM calculations are the most accurate while both RCM and GMC perform poorly in comparison.  Though the predictions of the CLM theorem are quite accurately predicted by all three numerical methods for high modulus contrast composites, the FEM results show the least error between the original and the translated effective properties while the RCM results show the largest error.  The exact relation for isotropic composites with components that have the same bulk moduli also shows that the FEM predictions are the most accurate though they are somewhat higher than the exact values.  However, no such distinction between the three methods can be made using Hill's equation.  These results reflect previous studies for high modulus contrast, high volume fraction polymer bonded explosives using FEM, RCM and RCM and show that exact relations can be used to determine the accuracy of numerical methods without performing detailed numerical studies.

  \section*{Acknowledgements}
  This research was supported by the University of Utah Center for the
  Simulation of Accidental Fires and Explosions (C-SAFE), funded by the     
  Department of Energy, Lawrence Livermore National Laboratory, under 
  subcontract B341493.

  \appendix
  \section*{Appendix}
  \setcounter{section}{1}
  %\appendix
%`\chapter{Plane Strain Stiffness and Compliance Matrices}\label{appA}
The components of the two-dimensional stiffness matrix can be computed from
two-dimensional plane strain finite element analyses.  However, the components
of the two-dimensional compliance matrix cannot be directly determined from
two-dimensional plane strain finite element analyses.  The reasons for these
are discussed in this appendix.  The approach taken to approximate the 
two-dimensional compliance matrix is also discussed.

  \subsection{Two-Dimensional Stiffness Matrix}
  The stress-strain relation for an anisotropic linear elastic material is given by
  \begin{equation}
    \left[\begin{array}{c} \Sigx \\ \Sigy \\ \Sigz \\ \Tauyz \\
	\Tauzx \\ \Tauxy \end{array}\right]  =
    \left[\begin{array}{cccccc} C_{11} & C_{12} & C_{13} & C_{14} & C_{15} & C_{16} \\
                    C_{12} & C_{22} & C_{23} & C_{24} & C_{25} & C_{26} \\
                    C_{13} & C_{23} & C_{33} & C_{34} & C_{35} & C_{36} \\
		    C_{14} & C_{24} & C_{34} & C_{44} & C_{45} & C_{46} \\
		    C_{15} & C_{25} & C_{35} & C_{45} & C_{55} & C_{56} \\
		    C_{16} & C_{26} & C_{36} & C_{46} & C_{56} & C_{66} \end{array}\right]
    \left[\begin{array}{c} \Epsx \\ \Epsy \\ \Epsz \\ 
	\Gamyz \\ \Gamzx \\ \Gamxy \end{array}\right] .
  \end{equation}
  For the plane strain assumption, we have,
  \begin{equation}
     \Epsz~=~\Gamyz~=~\Gamzx~=~0.
  \end{equation}
  Therefore, the stress-strain relation can be reduced to
  \begin{equation}
    \left[\begin{array}{c} \Sigx \\ \Sigy \\ \Tauxy \end{array}\right]  =
    \left[\begin{array}{ccc} C_{11} & C_{12} & C_{16} \\
                    C_{12} & C_{22} & C_{26} \\
                    C_{16} & C_{26} & C_{66} \end{array}\right]
    \left[\begin{array}{c} \Epsx \\ \Epsy \\ \Gamxy \end{array}\right] .
  \end{equation}
  The six terms in the apparent two-dimensional stiffness matrix reduce to
  four is the material is orthotropic, i.e.,
  \begin{equation}
    \left[\begin{array}{c} \Sigx \\ \Sigy \\ \Tauxy \end{array}\right]  =
    \left[\begin{array}{ccc} C_{11} & C_{12} & 0 \\
                    C_{12} & C_{22} & 0 \\
                    0 & 0 & C_{66} \end{array}\right]
    \left[\begin{array}{c} \Epsx \\ \Epsy \\ \Gamxy \end{array}\right] .
  \end{equation}
  The three constants $C_{11}$, $C_{12}$ and $C_{22}$ can be determined by
  the application of normal displacements in the '1' and '2' directions 
  respectively.  The constant $C_{66}$ can be determined using shear displacement
  boundary conditions in a finite element analysis.  Hence, it can be seen that
  the stiffness matrix can be calculated directly from two-dimensional plane
  strain based finite element analyses.  This is not true for the compliance 
  matrix.

  \subsection{Two-Dimensional Compliance Matrix}
  The strain-stress relation for an anisotropic linear elastic material can be
  written as
  \begin{equation}
    \left[\begin{array}{c} \Epsx \\ \Epsy \\ \Epsz \\ 
	\Gamyz \\ \Gamzx \\ \Gamxy \end{array}\right]  =
    \left[\begin{array}{cccccc} S_{11} & S_{12} & S_{13} & S_{14} & S_{15} & S_{16} \\
                    S_{12} & S_{22} & S_{23} & S_{24} & S_{25} & S_{26} \\
                    S_{13} & S_{23} & S_{33} & S_{34} & S_{35} & S_{36} \\
		    S_{14} & S_{24} & S_{34} & S_{44} & S_{45} & S_{46} \\
		    S_{15} & S_{25} & S_{35} & S_{45} & S_{55} & S_{56} \\
		    S_{16} & S_{26} & S_{36} & S_{46} & S_{56} & S_{66} \end{array}\right]
    \left[\begin{array}{c} \Sigx \\ \Sigy \\ \Sigz \\ \Tauyz \\
	\Tauzx \\ \Tauxy \end{array}\right] .
  \end{equation}
  The relationship between the stiffness matrix and the compliance matrix is
  \begin{equation}
    \left[\begin{array}{cccccc} S_{11} & S_{12} & S_{13} & S_{14} & S_{15} & S_{16} \\
                    S_{12} & S_{22} & S_{23} & S_{24} & S_{25} & S_{26} \\
                    S_{13} & S_{23} & S_{33} & S_{34} & S_{35} & S_{36} \\
		    S_{14} & S_{24} & S_{34} & S_{44} & S_{45} & S_{46} \\
		    S_{15} & S_{25} & S_{35} & S_{45} & S_{55} & S_{56} \\
		    S_{16} & S_{26} & S_{36} & S_{46} & S_{56} & S_{66} \end{array}\right]
     = 
    \left[\begin{array}{cccccc} C_{11} & C_{12} & C_{13} & C_{14} & C_{15} & C_{16} \\
                    C_{12} & C_{22} & C_{23} & C_{24} & C_{25} & C_{26} \\
                    C_{13} & C_{23} & C_{33} & C_{34} & C_{35} & C_{36} \\
		    C_{14} & C_{24} & C_{34} & C_{44} & C_{45} & C_{46} \\
		    C_{15} & C_{25} & C_{35} & C_{45} & C_{55} & C_{56} \\
		    C_{16} & C_{26} & C_{36} & C_{46} & C_{56} & C_{66} 
	\end{array}\right]^{-1}
  \end{equation}
  or, 
  \begin{equation}
     \mathbf{S} = \mathbf{C}^{-1}.
  \end{equation}
  It is obvious from the above equation that the apparent two-dimensional 
  compliance matrix is not equal to the inverse of the apparent two-dimensional 
  stiffness matrix, i.e., 
  \begin{equation}
    \left[\begin{array}{ccc} S_{11} & S_{12} & S_{16} \\
                    S_{12} & S_{22} & S_{26} \\
		    S_{16} & S_{26} & S_{66} \end{array}\right]  \neq
    \left[\begin{array}{ccc} C_{11} & C_{12} & C_{16} \\
                    C_{12} & C_{22} & C_{26} \\
		    C_{16} & C_{16} & C_{66} \end{array}\right]^{-1}.
  \end{equation}
  Hence, we cannot determine the two-dimensional compliance matrix if we only 
  know the two-dimensional stiffness matrix.
 
  Let us again examine the effect of the plane-strain assumption on the 
  stress-strain relation.  We then have
  \begin{equation}
    \left[\begin{array}{c} \Epsx \\ \Epsy \\ 0 \\\Gamxy 
	\end{array}\right]  =
    \left[\begin{array}{cccc} S_{11} & S_{12} & S_{13} & S_{16} \\
                    S_{12} & S_{22} & S_{23} & S_{26} \\
                    S_{13} & S_{23} & S_{33} & S_{36} \\
		    S_{16} & S_{26} & S_{36} & S_{66} \end{array}\right]
    \left[\begin{array}{c} \Sigx \\ \Sigy \\ \Sigz \\ 
	 \Tauxy \end{array}\right] .
  \end{equation}
  For orthotropic materials, this relation simplifies to
  \begin{equation}
    \left[\begin{array}{c} \Epsx \\ \Epsy \\ 0 \\ 
	\Gamxy \end{array}\right]  =
    \left[\begin{array}{cccc} S_{11} & S_{12} & S_{13} & 0 \\
                    S_{12} & S_{22} & S_{23} & 0 \\
                    S_{13} & S_{23} & S_{33} & 0 \\
		    0 & 0 & 0 & S_{66} \end{array}\right]
    \left[\begin{array}{c} \Sigx \\ \Sigy \\ \Sigz \\ 
	 \Tauxy \end{array}\right] .
  \end{equation}
  This equation shows that we need to know the stress $\Sigz$ to determine
  the terms of the compliance matrix and hence three-dimensional analyses are
  necessary.  If we assume plane stress, we can determine the terms of the 
  matrix $\mathbf{S}$ directly.  However, the apparent two-dimensional 
  compliance matrix for plane stress is not equal to that for plane strain and
  hence we cannot apply this method for our purposes.  This is why the plane
  strain compliance matrix cannot be determined using two-dimensional finite
  element analyses only.

  \subsection{Approximation of Compliance Matrix}
  The two-dimensional compliance matrix can be determined approximately 
  for materials with square symmetry by assuming that $S_{13}$, $S_{23}$ and 
  $S_{33}$ are known.  Let,
  \begin{equation}
     S_{13} = S_{23} = -\frac{\nu_3}{E_3} \\
     S_{33} = \frac{1}{E_3} 
  \end{equation}
  where, $\nu_3$ is the Poisson's ratio in the out-of-plane direction and 
  $E_3$ is the Young's ratio in that direction.  Then, for a material with
  square symmetry,
  \begin{equation}
    \left[\begin{array}{c} \Epsx \\ \Epsy \\ 0 \\
	\Gamxy \end{array}\right]  =
    \left[\begin{array}{cccc} S_{11} & S_{12} & -\frac{\nu_3}{E_3} & 0 \\
                    S_{12} & S_{11} & -\frac{\nu_3}{E_3} & 0 \\
                    -\frac{\nu_3}{E_3} & -\frac{\nu_3}{E_3} & \frac{1}{E_3} & 0 \\
		    0 & 0 & 0 & S_{66} \end{array}\right]
    \left[\begin{array}{c} \Sigx \\ \Sigy \\ \Sigz \\ 
	 \Tauxy \end{array}\right] .
  \end{equation}
  Inverting the relation, we have,
  \begin{equation}
    \left[\begin{array}{c} \Sigx \\ \Sigy \\ \Sigz \\ 
	 \Tauxy \end{array}\right]  =
    \left[\begin{array}{cccc} C_{11} & C_{12} & C_{13} & 0 \\
                    C_{12} & C_{11} & C_{23} & 0 \\
                    C_{13} & C_{23} & C_{33} & 0 \\
		    0 & 0 & 0 & C_{66} \end{array}\right]
    \left[\begin{array}{c} \Epsx \\ \Epsy \\ 0 \\
	\Gamxy \end{array}\right].
  \end{equation}
  \begin{eqnarray*}
    \text{where,}~&
      C_{11} = \frac{E_3 S_{11} - \nu_3^2}{E_3 S_{11}^2 - 2\nu_3^2 S_{11} 
		- E_3 S_{12}^2 + 2\nu^2 S_{12}}~,& \\
      & C_{12} = \frac{-E_3 S_{12} + \nu_3^2}{E_3 S_{11}^2 - 2\nu_3^2 S_{11} 
		- E_3 S_{12}^2 + 2\nu^2 S_{12}}. & 
  \end{eqnarray*}
  Note that it is not necessary to know $C_{13}$, $C_{23}$ and $C_{33}$ to determine
  $S_{11}$ and $S_{12}$.  

  We can write the above relations between $C_{11},C_{12}$ and $S_{11},S_{12}$ 
  in the form
  \begin{eqnarray}
    E_3 S_{11}^2 - \left(\frac{E_3}{C_{11}} + 2\nu_3^2\right) S_{11}
    - \left(E_3 S_{12}^2 - 2\nu_3^2 S_{12} - \frac{\nu_3^2}{C_{11}}\right) & = 0,\\
    E_3 S_{12}^2 - \left(\frac{E_3}{C_{12}} + 2\nu_3^2\right) S_{12}
    - \left(E_3 S_{11}^2 - 2\nu_3^2 S_{11} + \frac{\nu_3^2}{C_{12}}\right) & = 0.
  \end{eqnarray}
  In simplified form,
  \begin{eqnarray}
     A_1 S_{11}^2 + B_1 S_{11} + C_1 & = 0, \\
     A_2 S_{12}^2 + B_2 S_{12} + C_2 & = 0.
  \end{eqnarray}
  We can solve these quadratic equations to get expressions for $S_{11}$ and
  $S_{12}$ as
  \begin{eqnarray}
     S_{11} & = \frac{-B + \sqrt{B^2 - 4 A C}}{2A}, \\
     S_{12} & = \frac{-B - \sqrt{B^2 - 4 A C}}{2A}.
  \end{eqnarray}
  Knowing $C_{11}$, $C_{12}$, $E_3$ and $\nu_3$ these two equations can be solved
  iteratively to determine $S_{11}$ and $S_{12}$.  The values of $C_{11}$ and
  $C_{12}$ can be determined using the procedure outlined at the beginning of
  this section.  It remains to be discussed how $E_3$ and $\nu_3$ are to be
  determined.

  \subsection{Determination of $E_3$ and $\nu_3$}
  Two methods can be used to determine the values of $E_3$ and $\nu_3$ for our
  calculations.  The first method is to assume that the rule of mixtures is
  accurate enough to determine the effective properties in the '3' direction.
  Thus, if the volume fraction of the first component is $f_1$ and that of the
  second component is $f_2$, we have,
  \begin{eqnarray}
    E_3 & = f_1 E_1 + f_2 E_2, \\
    \nu_3 & = f_1 \nu_1 + f_2 \nu_2.
  \end{eqnarray}
  where $E_i$ and $\nu_i$ are the Young's modulus and the Poisson's ratio of the
  $i$th component.

  The other option is to use the values of $S_{13}$, $S_{23}$ and $S_{33}$ 
  obtained from GMC since these are also quite accurate for the out of plane
  direction.  Thus, we have,
  \begin{eqnarray}
    E_3 & = \frac{1}{S_{33}^{\text{GMC}}},\\
    \nu_3 & = -S_{13}^{\text{GMC}} E_3.
  \end{eqnarray}

  This is the procedure we have use to determine the effective compliance matrices 
  discussed in this paper.

  \bibliographystyle{unsrt}
  \bibliography{exactPaper.bib}

\begin{thebibliography}{1}

\bibitem{Helsing97}
J.~Helsing, G.~W. Milton, and A.~B. Movchan.
\newblock Duality relations, correspondences, and numerical results for planar
  elastic composites.
\newblock {\em J. Mech. Phys. Solids}, 45(4):565--590, 1997.

\bibitem{Milton97}
G.~W. Milton.
\newblock Composites : a myriad of microstructure independent relations.
\newblock In T.~Tatsumi, E.~Watanabe, and T.~Kambe, editors, {\em Theoretical
  and Applied Mechanics (Proc. XIX International Congress of Theoretical and
  Applied Mechanics, Kyoto, 1996)}, pages 443--459. Elsevier, Amsterdam, 1997.

\bibitem{Aboudi96_1}
J.~Aboudi.
\newblock Micromechanical analysis of composites by the method of cells -
  update.
\newblock {\em Appl. Mech. Rev}, 49(10):S83--S91, 1996.

\bibitem{Banerjee02c}
B.~Banerjee and D.~O. Adams.
\newblock On predicting the effective elastic properties of polymer bonded
  explosives using the recursive cell method.
\newblock 2002.

\bibitem{Milton02}
G.~W. Milton.
\newblock {\em Theory of Composites}.
\newblock Cambridge University Press, New York, 2002.

\bibitem{Cherk92}
A.~V. Cherkaev, K.~A. Lurie, and G.~W. Milton.
\newblock Invariant properties of the stress in plane elasticity and
  equivalence classes of composites.
\newblock {\em Proc. R. Soc. Lond. A}, 438(1904):519--529, 1992.

\bibitem{Hill64}
R.~Hill.
\newblock Theory of mechanical properties of fibre-strengthened materials: I.
  elastic behaviour.
\newblock {\em J. Mech. Phys. Solids}, 12:199--212, 1964.

\bibitem{Banerjee02a}
B.~Banerjee and D.~O. Adams.
\newblock Effective elastic moduli of polymer bonded explosives from finite
  element simulations.
\newblock 2002.

\bibitem{Banerjee02b}
B.~Banerjee and D.~O. Adams.
\newblock Application of the generalized method of cells to polymer bonded
  explosives.
\newblock 2002.

\end{thebibliography}

\end{document}